\documentclass[%
 aip,
 amsmath,amssymb,
 reprint,%
]{revtex4-1}

\usepackage{graphicx}
\usepackage{dcolumn}
\usepackage{bm}
\usepackage{hyperref}
\usepackage[utf8]{inputenc}
\usepackage[T1]{fontenc}
\usepackage{mathptmx}
\usepackage{amsmath}
\usepackage{etoolbox}
\usepackage{multirow}
\usepackage{parskip}

\makeatletter
\def\@email#1#2{%
 \endgroup
 \patchcmd{\titleblock@produce}
  {\frontmatter@RRAPformat}
  {\frontmatter@RRAPformat{\produce@RRAP{*#1\href{mailto:#2}{#2}}}\frontmatter@RRAPformat}
  {}{}
}%
\makeatother
\begin{document}


\title{Two dimensional LiMgAs; a novel Topological Quantum Catalyst for Hydrogen Evolution Reaction}
\author{Raghottam M. Sattigeri}
\author{Prafulla K. Jha}%
\email{prafullaj@yahoo.com}
\affiliation{Department of Physics, Faculty of Science, The Maharaja Sayajirao University of Baroda, Vadodara-390002, Gujarat, India}%

\author{Piotr \'Spiewak}%
\affiliation{Materials Design Division, Faculty of Materials Science and Engineering, Warsaw University of Technology, 141 Wo{\l}oska Str., 02-507 Warsaw, Poland}%

\author{Krzysztof J. Kurzyd{\l}owski}%
\affiliation{Faculty of Mechanical Engineering, Bia{\l}ystok University of Technology, 45C Wiejska Str., 15-351, Bia{\l}ystok, Poland}%
%


\begin{abstract}
Quantum materials such as Topological Insulators (TI) have been promising due to diverse applications of their robust
surface/edge states in the bulk (3D) and two-dimensional (2D) regime. Such conducting surface states in 3D systems,
host "\textit{electron bath}" which are known to facilitate catalysis. However, the analogous effects in 2D scenarios wherein, conducting helical edge states giving rise to Fermionic accumulation has been scarcely addressed. Using density functional theory based \textit{first-principles} calculations, we demonstrate that, the conducting edge states in 2D TI such as LiMgAs can be exploited to facilitate excellent catalytic response towards Hydrogen evolution reactions. The Gibbs free energy in such cases was found to be as low as $-$0.02 eV which is quite superior as compared to other materials reported in literature. The concept presented herein can be extended to other well known 2D TI and used to realise novel topological quantum catalysts for ultra-high performance and efficient catalytic applications.
\end{abstract}

\keywords{2D Topological Insulators, Topological Catalyst, DFT}

\maketitle


\begin{figure*}[ht]
	\includegraphics[width=15cm]{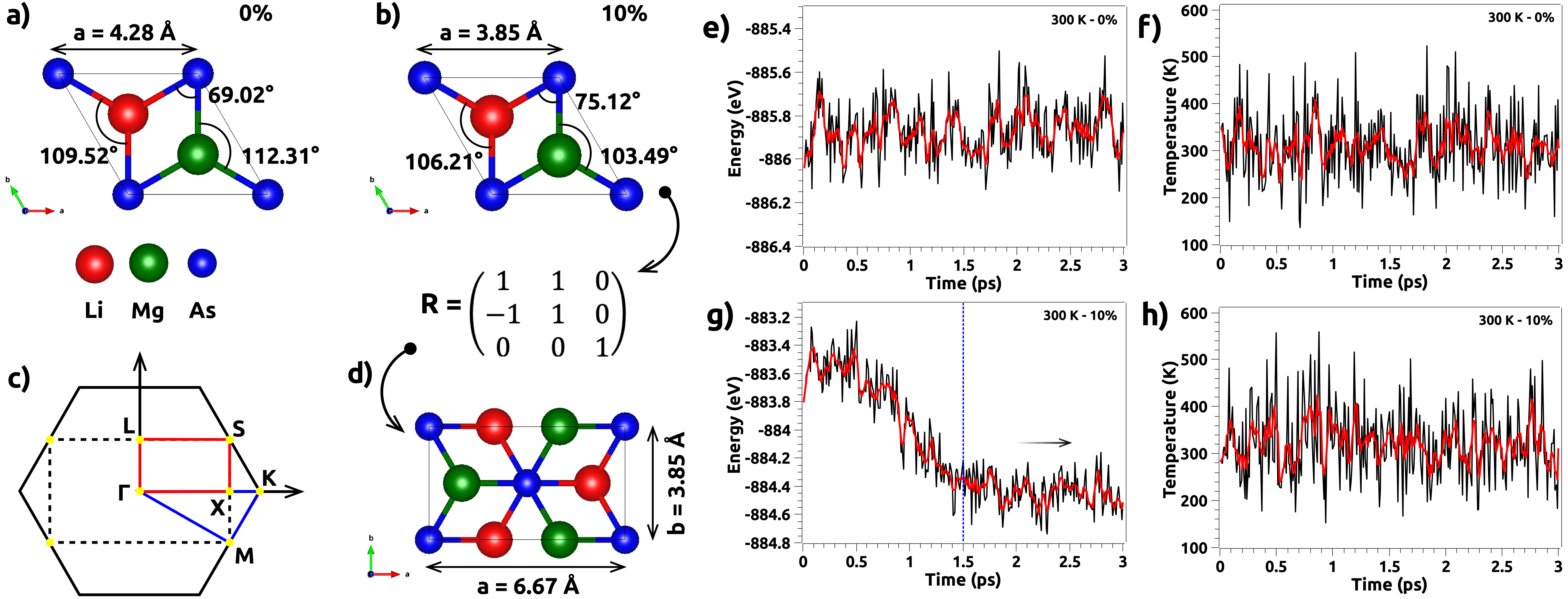}
	\caption{\label{fig1} (a,b) The unit cells of 2D LiMgAs indicating the structural parameters such as, lattice constant (\textit{a}) and the angle between atoms at, 0\% stress and 10\% biaxial compressive stress. The correspinding monolayer thickness (\textit{h}) is 1.60 \AA{} and 1.95 \AA{} respectively. (c) The hexagonal BZ of 2D LiMgAs superimposed with the orthorhombic BZ of LiMgAs obtained by using the rotation transformation matrix ($\mathnormal{\textbf{R}}$) and presented in (d). (e-h) Energy versus time and temperature versus time plots of the AIMD simulations of 2D LiMgAs under, pristine conditions (0\% strain) and at 10\% compressive stress for 2 picoseconds (3000 femtoseconds) with thermostat set at 300 K.}
	\vskip -1.5em
\end{figure*}

With increase in global energy demands, realising efficient and clean energy generation has attained great importance over the past few decades.\cite{1obama2017irreversible,2seh2017combining,3ding2019transforming} To this effect, generation of energy using Hydrogen-rich resources has proven to be effective with prominant use of Platinum group metals such as, Pt, Ru, Rh, Ir and Pd as a catalysts towards Hydrogen evolution reaction (HER).\cite{4zeng2015recent} However, the major issue is, the cost, rarity and short lifetime of electrodes associated with the Platinum group metals. This motivated search for other cost-effective and abundant alternatives. Till date, combination of metals (such as, Pt, Pd, Ir, Ru, Au, Ni), binary metal alloys, 3\textit{d} transition metal hydroxides, phosphides, carbides and transition metal chalcogenides, nitrides, borides, oxides and sulphides have been proposed as cost-effective alternatives.\cite{4zeng2015recent,5strmcnik2016design} However, their efficiency and performance are governed by the pH of surrounding media (with qualitative and quantitative differences in acidic and alkaline media).\cite{5strmcnik2016design} 

The major factor affecting the catalytic activity, efficiency and performance of such materials is surface states which arise from; [111] surfaces of noble metals and dangling bonds on semi-conductor surface significantly affecting the physio-chemical processes occuring on the surfaces.\cite{6chen2011co} This is due to the spatial proximity and orbital natures across the Fermi level. Such states are quite sensitive and not robust against; minor local deviations/variations (due to impurities and defects), surface deterioration (owing to redox reactions and corresponding degradation) altering the atomic termination and crystal orientations.\cite{6chen2011co} Apart from these factors as proposed by Gerischer and Parson,\cite{7sabatier1920catalyse,8gerischer1958mechanismus,9parsons1958rate,10quaino2014volcano} the requirement arising from the Sabatier's principle pertaining to the Gibb's free energy ($\Delta$G) i.e., thermoneutral condition with $\Delta$G $=$ 0 (yeilding optimal exchange current densities) is another major challenge. Realising materials which are capable of efficiently addressing the said physio-chemical problems has been the most challengning task from materials design point of view.

With this background, topological quantum materials (such as, topological insulators (TI), topological (Dirac, Weyl, nodal line) semimetals (TSM)) appear to be the most promising materials which can address the problems with existig catalysts.\cite{11xie2022progress} The character which makes them stand out from existing catalysts is, conducting surface/edge states and Fermi arcs (which are relatively robust against \textit{mild} impurities (magnetic/non-magnetic), surface oxidations and degradations) ensuring relatively longer life-times of catalysts and exchange of high mobility electrons facilitating HER.\cite{11xie2022progress} This approach was first proposed in 2011 wherein, CO and O$_2$ adsorptions over gold-covered TI Bi$_2$Se$_3$ surface were investigated; with emphasis on the persistance of surface states as long as the bulk gap is retained owing to the time-reversal symmetry.\cite{6chen2011co} It was expected that, such design strategy would establish alternatives to conventional \textit{d-}band theory of heterogeneous catalysis. However, ever since then, the interest in exploring topological quantum catalysts (TCat) made up of TI and TSM had almost quenched until, recently; when the field was revived with new perspectives for exploraing heterogenous catalysts based on, TSM, Chiral crystals, Full/Half-Heusler alloys and Perovskite manganites.\cite{12li2020heterogeneous}

TSM might be promising TCat but, in some scenarios such as, Dirac and Weyl semimetals, the weak electrostatic screening strength gives rise to lower carrier densities around the Fermi level making them less reliable.\cite{13wang2017quantum} This motivatd us to revisit and investigate heterogenous catalysis in two dimensional (2D) TI which are known to host high carrier mobilities, high carrier densities across Fermi level and robust charge accumulations along the edges. In this letter we report excellent catalytic properties of 2D TI LiMgAs investigated using Density Functional Theory (DFT). This system has been previously investigated by us for its TI nature which can be dimensionally engineered from the [111] crystal plane of bulk Half-Heusler (HH) compound LiMgAs.\cite{14sattigeri2021dimensional} Here, we explore two types of edge sites, (a) along the [100] plane and (b) along the [010] plane of 2D TI LiMgAs. We find that, the catalytic activity and performance is persistent along the edges and gets enhanced when spin-orbit coupling (SOC) is included. With this effort, we believe that, dimensionally engineered 2D TI from different planes of the HH compounds can be explored as candidate TCat. This approach has potential of discovering large number of materials since, several HH compounds have been predicted and realised experimentally till date owing to the permutations and combinations of the \textit{s$-$} and \textit{p$-$} block elements which typically make up HH compounds.\cite{15graf2011simple,16casper2012half,17yan2014half,18felser2015basics} TCat would thus translate as an alternative to the conventional \textit{d$-$}band theory and single atom catalysis for varied applications.

\begin{figure*}[ht]
	\includegraphics[width=15cm]{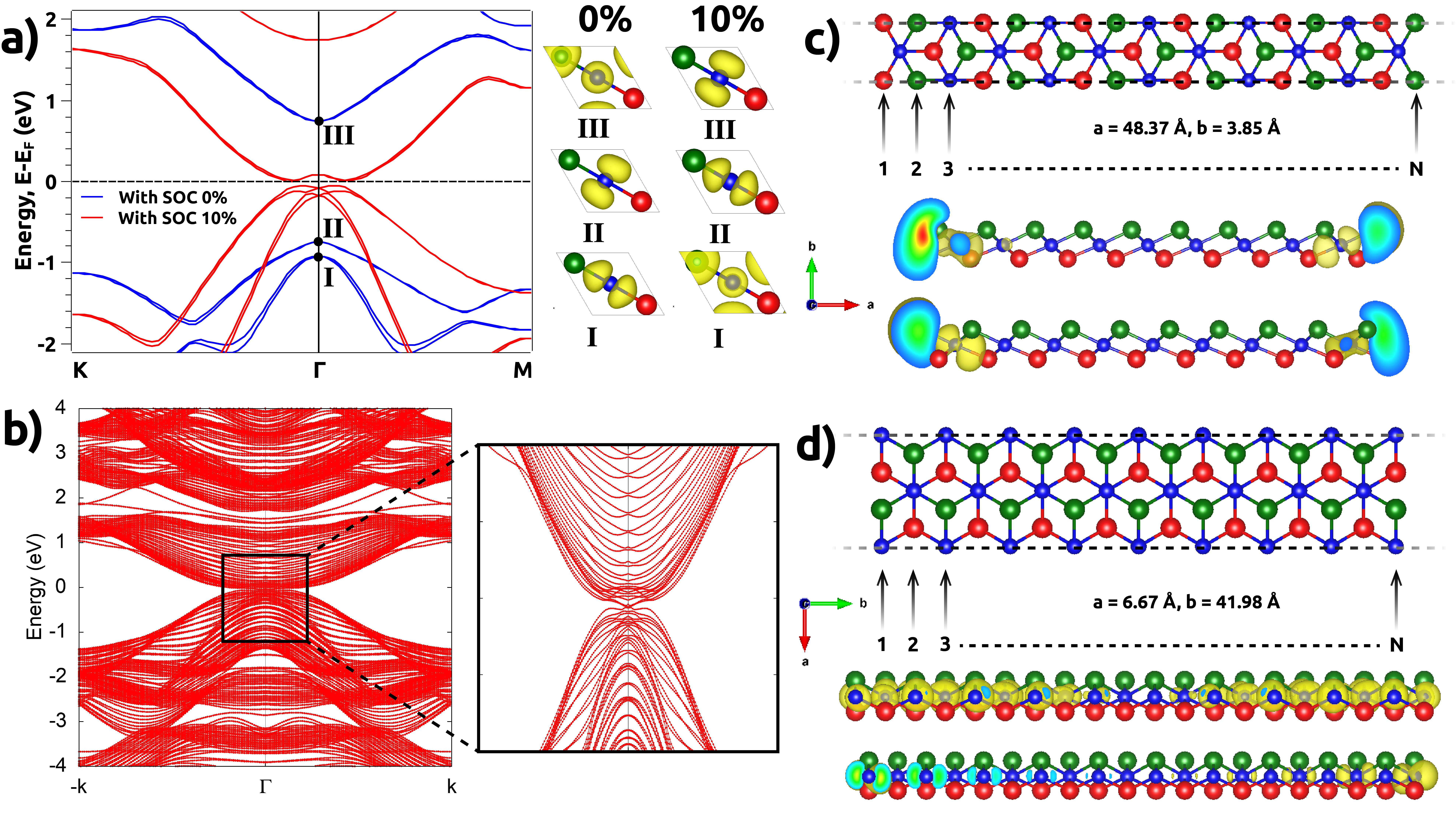}
	\caption{\label{fig2} (a) Electronic band structure with SOC at 0\% and 10\% biaxial compressive stress alongside the charge densities of bands I, II and III. (b) Slab band structure indicating the semi-metallic edge sites. (c,d) Zig-zag (periodic along [010] crystal direction) and Planar (periodic along [100] crystal direction) terminations of the nanoribbons (alongwith lattice parameters) and their corresponding edge Fermionic accumulations.}
	\vskip -1.5em
\end{figure*}

We performed Density Functional Theory based \textit{first-principles} calculations using Quantum ESPRESSO.\cite{19giannozzi2009quantum} For calculations without SOC, we use scalar-relativistic and norm-conserving Martins-Troullier pseudopotentials and for calculations with SOC, we use fully-relativistic Projector Augemented Wave (PAW) pseudopotentials. The generalized gradient approximation was implemented with Perdew-Burke-Ernzerhof type of exchange-correlation functional.\cite{20perdew1996generalized,21blochl1994projector} We used the hexagonal unit cell from the previous work and transformed it into orthorhombic unit cell using the rotation matrix ($\mathnormal{\textbf{R}}$) which transforms the hexagonal basis vector into orthorhombic basis vector (as evident from Fig. \ref{fig1}). This orthorhombic unit cell was then transformed into nanoribbons (with ribbon width N = 15 to avoid edge-edge interactions) which are periodic along, (a) [100] and (b) [010] crystal directions with zig-zag and planar like edges respectively (presented in Fig. \ref{fig2}(c,d)). The corresponding uniform momentum Monkhorst-Pack grid (\textit{k}-mesh) for brillouin zone (BZ) sampling was set to be 1 $\times$ 6 $\times$ 1 and 6 $\times$ 1 $\times$ 1 respectively.\cite{22monkhorst1976special} A vacuum of 25 \AA{} along the [001] direction and 17 \AA{} along the aperiodic directions in the nanoribbon configurations was imposed to avoid interactions due to periodic images. We also performed the calculations with 3 $\times$ 3 $\times$ 1 supercell of 2D LiMgAs (with 25 \AA{} vacuum along the [001] crystal direction) with \textit{k}-mesh for BZ sampling as 6 $\times$ 6 $\times$ 1. The optimized kinetic energy cutoff used in all the calculations was set to 60 Ry and the systems were relaxed with force convergence criteria of $<$ 10$^{-6}$ a.u. Throughout the calculations, Grimme correlations were incorporated to address the van der Waals effects.\cite{23grimme2006semiempirical} From the perspective of stability and room temperature viability, we performed ab initio molecular dynamics simulations (AIMD) for 3 picoseconds (3000 femtoseconds) with thermostat set at 300 K. We calculate the slab band structures using WannierTools wherein the tight-binding Hamiltonian generated by Wannier90 is implemented.\cite{24wu2018wanniertools,25mostofi2008wannier90}



The work presented in this letter builds up on 2D TI LiMgAs which is designed dimensionally and exhibits non-trivial topology engineered by application of biaxial stress. Under pristine conditions (without any pressure), 2D LiMgAs is characterised by a structure similar to 1T-MoS$_2$ with, As atoms forming a hexagonal lattice occupying the octahedral coordinations between two layers of hexagonal-closed-packed structures made up of Li and Mg (as presented in Fig. \ref{fig1}(a)). Such arragnement is governed by the P$\overline{3}$m1 space group and the octahedral belongs to the D$_{3d}$ tetragonal symmetry.\cite{26jayabal2018metallic} At around 10\% biaxial compressive stress the structural intigrity is retained (as evident from Fig. \ref{fig1}(b)) however, the system undergoes a quantum topological phase transition from a trivial insulator to a non-trivial TI. We assess the stability and viability of these systems at room temperature (300 K) by performing AIMD calculations. As evident from Fig. \ref{fig1}(e) the pristine structure of LiMgAs (at 0\% stress) thermalises and is stable throughout the simulation. However, the system at 10\% biaxial compressive stress has large fluctuations in energy (as evident from Fig. \ref{fig1}(g)) initially but, after 1.5 picoseconds (highlighted by blue line in Fig. \ref{fig1}(g)) of simulations, the system thermalises; attaining structural stability which is evident from small energy fluctuations thereafter. The corresponding thermostat variations in both the cases are also presented in Fig. \ref{fig1}(f,h).

\begin{figure*}[ht]
	\includegraphics[width=14cm]{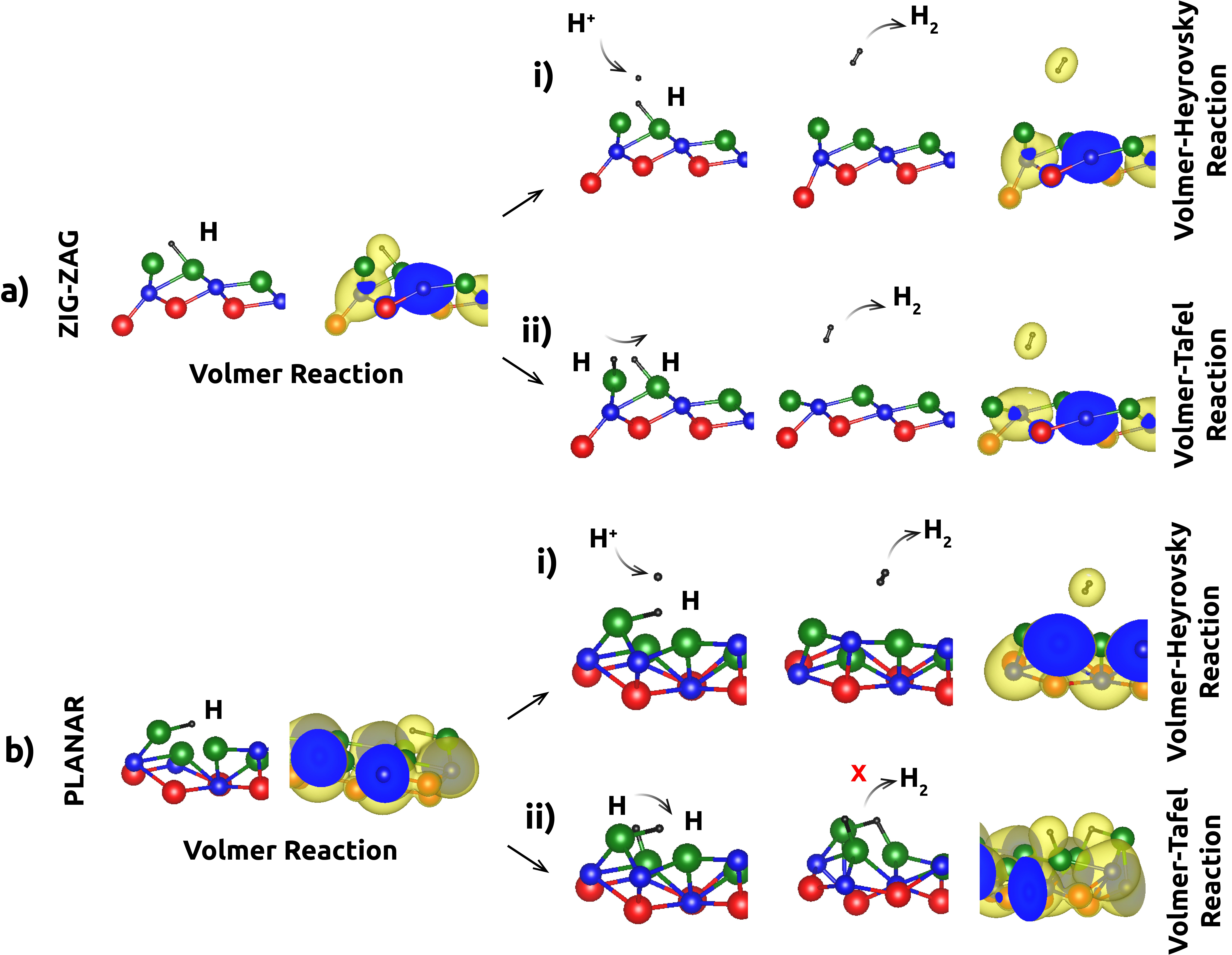}
	\caption{\label{fig3} HER mechanism involving the primary Volmer reaction and the secondary reactions (Volmer-Heyrovsky and Volmer-Tafel) along the edge sites of, (a) Zig-zag and (b) Planar terminations of the LiMgAs nanoribbon.}
	\vskip -1.5em
\end{figure*}

We briefly discuss the topological properties of 2D LiMgAs. Figure \ref{fig2}(a) presents the electronic band structure of 2D LiMgAs at 0\% and 10\% biaxial compressive stress under the SOC regime. Apart from the s$-$, p$-$ orbital inversion mechanism which has been previously discussed, we observe exchange of charge densities across the Fermi level. This exchange is clearly evident from the charge densities of bands I, II and III at 0\% and 10\% biaxial compressive stress which reafirms the non-trivial topological phase transitions. At 0\% stress, the band gap of the system under SOC is 1.49 eV whereas, upon quantum phase transition (at 10\% stress) it turns out to be 0.16 eV which indicates that 2D LiMgAs exhibits a non-trivial large-gap which is superior to several materials.\cite{27zhao2020two} This implies that, the edge states would be robust even at room temperature. Also due to, high velocity of Fermions, 2D LiMgAs crops up as an interesting candidate for catalytic applications. Figure \ref{fig2}(b) exhibits the slab band structure calculated using the tight-binding Hamiltonian model. It is clear that, the edge sites host a non-trivial semi-metallic character while the band structure hosts a SOC induced gap at 10\% biaxial compressive stress. 

With the help of rotation matrix ($\mathnormal{\textbf{R}}$) we transform the hexagonal unit cell into an orthorhombic unit cell at 10\% biaxial compressive stress (as presented in Fig. \ref{fig1}(b,d)). Using this orthorhombic phase, we design two nanoribbon configurations with, (a) zig-zag and (b) planar like edge terminations (by exploiting the periodicity along [100] and [010] crystal direction). The nanoribbon configurations presented in Fig. \ref{fig2}(c,d) have formation energies E$_f$ ($=$ E$_{\textit{LiMgAs}} -$ E$_{\textit{Li}} -$ E$_{\textit{Mg}} -$ E$_{\textit{As}}$, where E$_{\textit{LiMgAs}}$ is the energy of entire system and E$_{\textit{X}}$ with \textit{X} $=$ Li, Mg, As; is the energy of constituent elements) $-$2.03 eV/atom and $-$2.90 eV/atom respectively. These energies are relatively lower as compared to their 2D counterpart which has formation energy $-$1.70 eV/atom, indicating towards stability and high probability of formation of nanoribbon configurations. It is also evident from these configurations (presented in Fig. \ref{fig2}(c,d)) that, the edge sites have robust Fermionic accumulations owing to the non-trivial topology. These charactertic properties of 2D TI LiMgAs indicate that, it can potentially host TCat along the edge sites as compared to the basal sites.

In order to confirm the superior activity along the edge sites as compared to the basal sites in heterogenous catalysis such as HER and to affirm TCat, we begin our calculations on a 3 $\times$ 3 $\times$ 1 supercell of the hexagonal 2D TI LiMgAs which exposes the basal sites for HER. We begin our investigations with the primary Volmer reaction on different sites such as, top (of Li, Mg, As) and bridge sites (such as, Li-Mg, Mg-As, Li-As). Governed by Sabatier's principle, the interaction of Hydrogen (H) atom during the primary reaction should be optimal (i.e., the atom should interact neither strongly nor weakly with the catalytic surface). This implies that during the primary reaction (under neutral pH and reduction potential U $=$ 0), the parameters such as, adsorption energy ($\Delta$E$_{H}^{ads} =$ E$_{LiMgAs+H} -$ E$_{LiMgAs} - \frac{1}{2}$E$_{H_2}$) where, E$_{LiMgAs+H}$ is the total energy of H adsorbed LiMgAs, E$_{\textit{LiMgAs}}$ is the total energy of LiMgAs and E$_{H_2}$ is the total energy of H$_2$ molecule) and Gibbs free energy ($\Delta$G$_{H} = \Delta$E$_{H}^{ads} + \Delta_{ZPE} -$ T$\Delta$S where, $\Delta_{ZPE}$ is the change in the zero-point energy of adsorbed and gaseous state of H and $\Delta$S is the corresponding change in the entropy and T is room temperature) should ideally be 0.24 eV and 0 eV respectively.\cite{28lee2017theoretical} However, in our case, during the primary reaction, the As site was found to be the most active site amongst all the basal sites with adsorption and Gibbs free energy without SOC (with SOC) as $-$2.38 eV ($-$2.46 eV) and $-$2.14 eV ($-$2.22 eV) respectively (see Supplementary Information (SI), Tab. 1 for other sites). This implies that the basal site strongly interacts with the adsorbed H indicating that the basal sites wont be ideal catalysts.



In sharp contrast to the interactions at basal sites, the edge sites exhibit excellent activity during the primary reaction due to robust Fermionic accumulations. The different reaction mechanisms investigated herein are presented in Fig. \ref{fig3}(a,b) for both zig-zag and planar termination configurations of nanoribbons. The sites investigated for HER were, top (Li, Mg, As) and bridge sites (Li-As, Mg-As). During primary reaction, the most active sites in the zig-zag and planar configurations without SOC (with SOC) are, As (As and Li-As) and As (As, Li) respectively. The most ideal adsorption and Gibbs free energy in the zig-zag configuration (under SOC) was found to be, $-$0.264 eV and $-$0.024 eV at As top site and, $-$0.26 eV and $-$0.02 eV at Li-As bridge site respectively (see SI, Tab. 2, for other sites). Similarly, for planar configuration the most ideal adsorption and Gibbs free energy (under SOC) was found to be, $-$0.213 eV and 0.026 eV at Li top site and, $-$0.212 eV and 0.027 eV at As top site respectively (see SI, Tab. 4, for other sites). To further assure and associate the origin of such excellent catalytic activity to the topological edge states, we performed calculations for primary reaction at the mid site of nanoribbons, specifically, on top of those sites which exhibit minimum Gibbs free energy under SOC (along the edges). We found that, in both the configurations i.e., zig-zag and planar, the interaction was of repelling nature with adsorption energy 0.118 eV and 0.679 eV respectively, while the corresponding Gibbs free energy were, 0.358 eV and 0.919 eV respectively. This firmly established the fact that, the superior catalytic behaviour indeed originates from the topologically robust edge sites.

We now proceed with the secondary reaction mechanisms which are Volmer-Heyrovsky (the H adsorbed in primary reaction interacts with a H$^+$ ion and evolves into H$_2$) and Volmer-Tafel (the H adsorbed in primary reaction interacts with a neighbouring H and evolves into H$_2$) along the optimal sites in both the zig-zag and planar configurations of the nanoribbons as presented in Fig. \ref{fig3}(a(i,ii), b(i,ii)) respectively. The adsorption and Gibbs free energy fomalism for Volmer-Heyrovsky intermediate reaction is slightly modified and given by, $\Delta$E$_{H^*}^{ads}(n) (= (\frac{1}{n}$ E$_{LiMgAs+nH^*} -$ E$_{LiMgAs} -$ $\frac{n}{2}$E$_{H_2}$) and $\Delta$G$_{H^*}(n) (= \Delta$E$_{H^*}^{ads}(n) + \Delta_{ZPE} -$ T$\Delta$S) respectively with \textit{n} $=$ 2 and terms having their usual meaning.\cite{28lee2017theoretical} For Volmer-Tafel intermediate reactions, we follow the modified formalism proposed by Lee et.al.\cite{28lee2017theoretical} 

\begin{figure}[ht]
	\includegraphics[width=8cm]{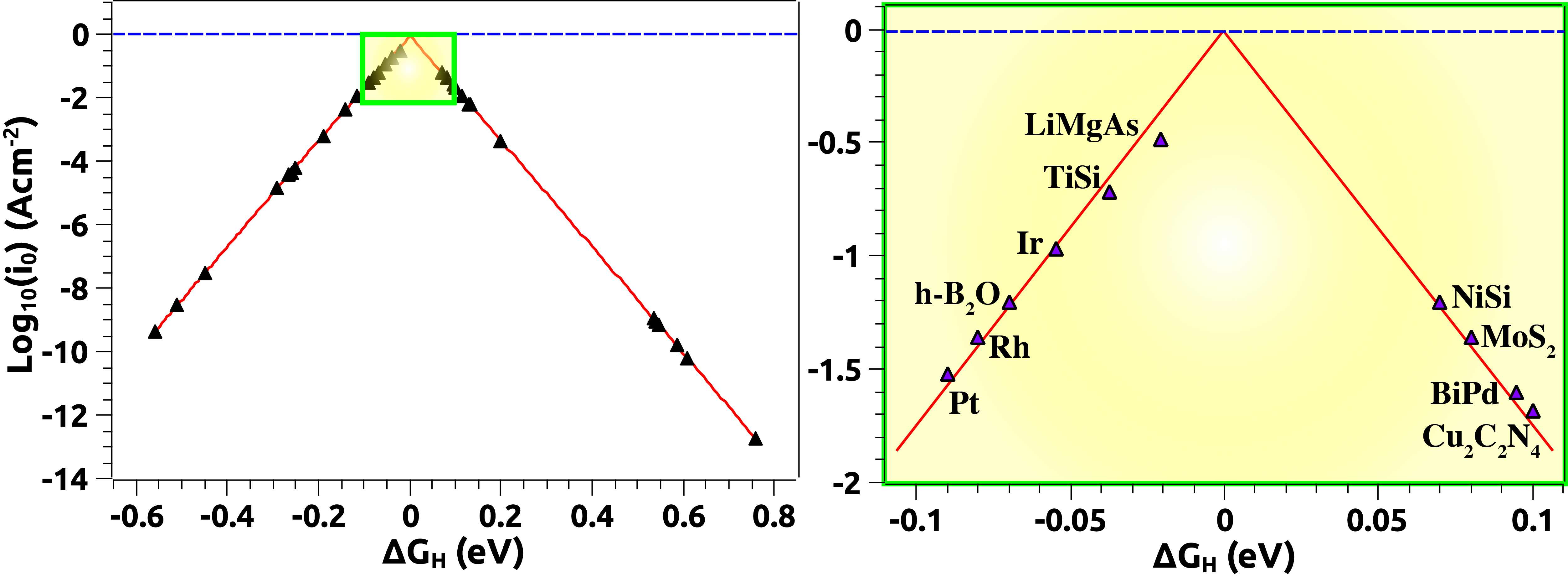}
	\caption{\label{fig4} Volcano plot generated here is based on our recomputations using, exchange current densities (for rate constant \textit{k}$_0 =$ 1 s$^{-1}$ site$^{-1}$, according to N{\o}rskov's assumptions)\cite{29norskov2005trends} and Gibbs free energy data from literature for various materials (see SI, Tab. 6 for details).\cite{2seh2017combining,30Pt-wang2017precise,31Pd-SnTe-qu2021highly,32h-B2O-zhao2020strain,33NiSi-liu2022theoretical,34MoS2-hinnemann2005biomimetic,35TiSi-li2018topological,36BiPd-he2019topological,37Ni-VAl3-kong2021development,38W-PdGa-yang2021transition,39vol-plot2-ekspong2020hydrogen,40PtGa-PtAl-yang2020topological,41Cu2C2N4-wang2021topological,29norskov2005trends} It is evident that, LiMgAs outperforms other topological materials placing it closer to the Sabatiers optimum (highlighted in yellow).}
	\vskip -1em
\end{figure}

We find that, the zig-zag configuration facilitates both the secondary reactions Volmer-Heyrovsky and, Volmer-Tafel (as evident from Fig. \ref{fig3}(a(i,ii))) based on the adsorption and Gibbs free energy which are, $-$0.201 eV and 0.03 eV and, $-$0.296 eV and $-$0.05 eV respectively. Based on the Gibbs free energy, it is evident that, the Volmer-Heyrovsky and Volmer-Tafel mechanisms are exothermic and endothermic in nature respectively. However, the planar configuration only facilitates the Volmer-Heyrovsky reaction along the optimal sites As and Li (as evident from Fig. \ref{fig3}(b(i,ii))) with similar adsorption and Gibbs free energy of $-$0.53 eV and $-$0.29 eV respectively. Hence, the zig-zag configuration is highly sensitive towards HER as compared to the planar configuration. Overall, we observe that, the edge sites favour HER rather than basal sites. Such superior activity towards HER places 2D TI LiMgAs at the top of volcano plot closer to Sabatier's optimum (as evident from Fig. \ref{fig4}) with Gibbs free energy as low as $-$0.02 eV; giving rise to exchange current density (Log$_{10}(i_0)$) of the order of $-$0.5 Acm$^{-2}$. Also, the positive Gibbs free energy at certain sites (see SI) thermodynamically favours and readily facilitates HER. These features make 2D TI LiMgAs a \textit{novel} candidate for catalytic applications.

In summary, we propose that, 2D topological materials such as 2D TI can be novel quantum catalysts owing to their non-trivial and robust topologies (providing high density charge accumulation and carrier mobility) alongside high surface-to-volume ratio. In this letter we discuss the case of 2D TI LiMgAs which exhibits excellent catalytic activity towards HER. We propose that, multilayered structures of such 2D materials can be designed to create surfaces with such excellent catalytic properties. Also, we suggest that, large-gap TI would perform better at room temperature due to the persistence of edge states at such conditions.

\small 
\section*{REFERENCES}
\vskip -1.5em
\nocite{*}
\bibliography{aipsamp}

\providecommand{\noopsort}[1]{}\providecommand{\singleletter}[1]{#1}%
\begin{thebibliography}{41}%
\makeatletter
\providecommand \@ifxundefined [1]{%
 \@ifx{#1\undefined}
}%
\providecommand \@ifnum [1]{%
 \ifnum #1\expandafter \@firstoftwo
 \else \expandafter \@secondoftwo
 \fi
}%
\providecommand \@ifx [1]{%
 \ifx #1\expandafter \@firstoftwo
 \else \expandafter \@secondoftwo
 \fi
}%
\providecommand \natexlab [1]{#1}%
\providecommand \enquote  [1]{``#1''}%
\providecommand \bibnamefont  [1]{#1}%
\providecommand \bibfnamefont [1]{#1}%
\providecommand \citenamefont [1]{#1}%
\providecommand \href@noop [0]{\@secondoftwo}%
\providecommand \href [0]{\begingroup \@sanitize@url \@href}%
\providecommand \@href[1]{\@@startlink{#1}\@@href}%
\providecommand \@@href[1]{\endgroup#1\@@endlink}%
\providecommand \@sanitize@url [0]{\catcode `\\12\catcode `\$12\catcode
  `\&12\catcode `\#12\catcode `\^12\catcode `\_12\catcode `\%12\relax}%
\providecommand \@@startlink[1]{}%
\providecommand \@@endlink[0]{}%
\providecommand \url  [0]{\begingroup\@sanitize@url \@url }%
\providecommand \@url [1]{\endgroup\@href {#1}{\urlprefix }}%
\providecommand \urlprefix  [0]{URL }%
\providecommand \Eprint [0]{\href }%
\providecommand \doibase [0]{http://dx.doi.org/}%
\providecommand \selectlanguage [0]{\@gobble}%
\providecommand \bibinfo  [0]{\@secondoftwo}%
\providecommand \bibfield  [0]{\@secondoftwo}%
\providecommand \translation [1]{[#1]}%
\providecommand \BibitemOpen [0]{}%
\providecommand \bibitemStop [0]{}%
\providecommand \bibitemNoStop [0]{.\EOS\space}%
\providecommand \EOS [0]{\spacefactor3000\relax}%
\providecommand \BibitemShut  [1]{\csname bibitem#1\endcsname}%
\let\auto@bib@innerbib\@empty
\bibitem [{\citenamefont {Obama}(2017)}]{1obama2017irreversible}%
  \BibitemOpen
  \bibfield  {author} {\bibinfo {author} {\bibfnamefont {B.}~\bibnamefont
  {Obama}},\ }\href@noop {} {\bibfield  {journal} {\bibinfo  {journal}
  {Science}\ }\textbf {\bibinfo {volume} {355}},\ \bibinfo {pages} {126--129}
  (\bibinfo {year} {2017})}\BibitemShut {NoStop}%
\bibitem [{\citenamefont {Seh}\ \emph {et~al.}(2017)\citenamefont {Seh},
  \citenamefont {Kibsgaard}, \citenamefont {Dickens}, \citenamefont
  {Chorkendorff}, \citenamefont {N{\o}rskov},\ and\ \citenamefont
  {Jaramillo}}]{2seh2017combining}%
  \BibitemOpen
  \bibfield  {author} {\bibinfo {author} {\bibfnamefont {Z.~W.}\ \bibnamefont
  {Seh}}, \bibinfo {author} {\bibfnamefont {J.}~\bibnamefont {Kibsgaard}},
  \bibinfo {author} {\bibfnamefont {C.~F.}\ \bibnamefont {Dickens}}, \bibinfo
  {author} {\bibfnamefont {I.}~\bibnamefont {Chorkendorff}}, \bibinfo {author}
  {\bibfnamefont {J.~K.}\ \bibnamefont {N{\o}rskov}}, \ and\ \bibinfo {author}
  {\bibfnamefont {T.~F.}\ \bibnamefont {Jaramillo}},\ }\href@noop {} {\bibfield
   {journal} {\bibinfo  {journal} {Science}\ }\textbf {\bibinfo {volume}
  {355}},\ \bibinfo {pages} {eaad4998} (\bibinfo {year} {2017})}\BibitemShut
  {NoStop}%
\bibitem [{\citenamefont {Ding}\ \emph {et~al.}(2019)\citenamefont {Ding},
  \citenamefont {H{\"u}lsey}, \citenamefont {P{\'e}rez-Ram{\'\i}rez},\ and\
  \citenamefont {Yan}}]{3ding2019transforming}%
  \BibitemOpen
  \bibfield  {author} {\bibinfo {author} {\bibfnamefont {S.}~\bibnamefont
  {Ding}}, \bibinfo {author} {\bibfnamefont {M.~J.}\ \bibnamefont
  {H{\"u}lsey}}, \bibinfo {author} {\bibfnamefont {J.}~\bibnamefont
  {P{\'e}rez-Ram{\'\i}rez}}, \ and\ \bibinfo {author} {\bibfnamefont
  {N.}~\bibnamefont {Yan}},\ }\href@noop {} {\bibfield  {journal} {\bibinfo
  {journal} {Joule}\ }\textbf {\bibinfo {volume} {3}},\ \bibinfo {pages}
  {2897--2929} (\bibinfo {year} {2019})}\BibitemShut {NoStop}%
\bibitem [{\citenamefont {Zeng}\ and\ \citenamefont
  {Li}(2015)}]{4zeng2015recent}%
  \BibitemOpen
  \bibfield  {author} {\bibinfo {author} {\bibfnamefont {M.}~\bibnamefont
  {Zeng}}\ and\ \bibinfo {author} {\bibfnamefont {Y.}~\bibnamefont {Li}},\
  }\href@noop {} {\bibfield  {journal} {\bibinfo  {journal} {Journal of
  Materials Chemistry A}\ }\textbf {\bibinfo {volume} {3}},\ \bibinfo {pages}
  {14942--14962} (\bibinfo {year} {2015})}\BibitemShut {NoStop}%
\bibitem [{\citenamefont {Strmcnik}\ \emph {et~al.}(2016)\citenamefont
  {Strmcnik}, \citenamefont {Lopes}, \citenamefont {Genorio}, \citenamefont
  {Stamenkovic},\ and\ \citenamefont {Markovic}}]{5strmcnik2016design}%
  \BibitemOpen
  \bibfield  {author} {\bibinfo {author} {\bibfnamefont {D.}~\bibnamefont
  {Strmcnik}}, \bibinfo {author} {\bibfnamefont {P.~P.}\ \bibnamefont {Lopes}},
  \bibinfo {author} {\bibfnamefont {B.}~\bibnamefont {Genorio}}, \bibinfo
  {author} {\bibfnamefont {V.~R.}\ \bibnamefont {Stamenkovic}}, \ and\ \bibinfo
  {author} {\bibfnamefont {N.~M.}\ \bibnamefont {Markovic}},\ }\href@noop {}
  {\bibfield  {journal} {\bibinfo  {journal} {Nano Energy}\ }\textbf {\bibinfo
  {volume} {29}},\ \bibinfo {pages} {29--36} (\bibinfo {year}
  {2016})}\BibitemShut {NoStop}%
\bibitem [{\citenamefont {Chen}\ \emph {et~al.}(2011)\citenamefont {Chen},
  \citenamefont {Zhu}, \citenamefont {Xiao},\ and\ \citenamefont
  {Zhang}}]{6chen2011co}%
  \BibitemOpen
  \bibfield  {author} {\bibinfo {author} {\bibfnamefont {H.}~\bibnamefont
  {Chen}}, \bibinfo {author} {\bibfnamefont {W.}~\bibnamefont {Zhu}}, \bibinfo
  {author} {\bibfnamefont {D.}~\bibnamefont {Xiao}}, \ and\ \bibinfo {author}
  {\bibfnamefont {Z.}~\bibnamefont {Zhang}},\ }\href@noop {} {\bibfield
  {journal} {\bibinfo  {journal} {Physical review letters}\ }\textbf {\bibinfo
  {volume} {107}},\ \bibinfo {pages} {056804} (\bibinfo {year}
  {2011})}\BibitemShut {NoStop}%
\bibitem [{\citenamefont {Sabatier}(1920)}]{7sabatier1920catalyse}%
  \BibitemOpen
  \bibfield  {author} {\bibinfo {author} {\bibfnamefont {P.}~\bibnamefont
  {Sabatier}},\ }\href@noop {} {} (\bibinfo {year} {1920})\BibitemShut
  {NoStop}%
\bibitem [{\citenamefont {Gerischer}(1958)}]{8gerischer1958mechanismus}%
  \BibitemOpen
  \bibfield  {author} {\bibinfo {author} {\bibfnamefont {H.}~\bibnamefont
  {Gerischer}},\ }\href@noop {} {\bibfield  {journal} {\bibinfo  {journal}
  {Bulletin des Soci{\'e}t{\'e}s Chimiques Belges}\ }\textbf {\bibinfo {volume}
  {67}},\ \bibinfo {pages} {506--527} (\bibinfo {year} {1958})}\BibitemShut
  {NoStop}%
\bibitem [{\citenamefont {Parsons}(1958)}]{9parsons1958rate}%
  \BibitemOpen
  \bibfield  {author} {\bibinfo {author} {\bibfnamefont {R.}~\bibnamefont
  {Parsons}},\ }\href@noop {} {\bibfield  {journal} {\bibinfo  {journal}
  {Transactions of the Faraday Society}\ }\textbf {\bibinfo {volume} {54}},\
  \bibinfo {pages} {1053--1063} (\bibinfo {year} {1958})}\BibitemShut {NoStop}%
\bibitem [{\citenamefont {Quaino}\ \emph {et~al.}(2014)\citenamefont {Quaino},
  \citenamefont {Juarez}, \citenamefont {Santos},\ and\ \citenamefont
  {Schmickler}}]{10quaino2014volcano}%
  \BibitemOpen
  \bibfield  {author} {\bibinfo {author} {\bibfnamefont {P.}~\bibnamefont
  {Quaino}}, \bibinfo {author} {\bibfnamefont {F.}~\bibnamefont {Juarez}},
  \bibinfo {author} {\bibfnamefont {E.}~\bibnamefont {Santos}}, \ and\ \bibinfo
  {author} {\bibfnamefont {W.}~\bibnamefont {Schmickler}},\ }\href@noop {}
  {\bibfield  {journal} {\bibinfo  {journal} {Beilstein journal of
  nanotechnology}\ }\textbf {\bibinfo {volume} {5}},\ \bibinfo {pages}
  {846--854} (\bibinfo {year} {2014})}\BibitemShut {NoStop}%
\bibitem [{\citenamefont {Xie}\ \emph {et~al.}(2022)\citenamefont {Xie},
  \citenamefont {Zhang}, \citenamefont {Weng},\ and\ \citenamefont
  {Chai}}]{11xie2022progress}%
  \BibitemOpen
  \bibfield  {author} {\bibinfo {author} {\bibfnamefont {R.}~\bibnamefont
  {Xie}}, \bibinfo {author} {\bibfnamefont {T.}~\bibnamefont {Zhang}}, \bibinfo
  {author} {\bibfnamefont {H.}~\bibnamefont {Weng}}, \ and\ \bibinfo {author}
  {\bibfnamefont {G.-L.}\ \bibnamefont {Chai}},\ }\href@noop {} {\bibfield
  {journal} {\bibinfo  {journal} {Small Science}\ ,\ \bibinfo {pages}
  {2100106}} (\bibinfo {year} {2022})}\BibitemShut {NoStop}%
\bibitem [{\citenamefont {Li}\ and\ \citenamefont
  {Felser}(2020)}]{12li2020heterogeneous}%
  \BibitemOpen
  \bibfield  {author} {\bibinfo {author} {\bibfnamefont {G.}~\bibnamefont
  {Li}}\ and\ \bibinfo {author} {\bibfnamefont {C.}~\bibnamefont {Felser}},\
  }\href@noop {} {\bibfield  {journal} {\bibinfo  {journal} {Applied Physics
  Letters}\ }\textbf {\bibinfo {volume} {116}},\ \bibinfo {pages} {070501}
  (\bibinfo {year} {2020})}\BibitemShut {NoStop}%
\bibitem [{\citenamefont {Wang}\ \emph
  {et~al.}(2017{\natexlab{a}})\citenamefont {Wang}, \citenamefont {Lin},
  \citenamefont {Wang}, \citenamefont {Yu},\ and\ \citenamefont
  {Liao}}]{13wang2017quantum}%
  \BibitemOpen
  \bibfield  {author} {\bibinfo {author} {\bibfnamefont {S.}~\bibnamefont
  {Wang}}, \bibinfo {author} {\bibfnamefont {B.-C.}\ \bibnamefont {Lin}},
  \bibinfo {author} {\bibfnamefont {A.-Q.}\ \bibnamefont {Wang}}, \bibinfo
  {author} {\bibfnamefont {D.-P.}\ \bibnamefont {Yu}}, \ and\ \bibinfo {author}
  {\bibfnamefont {Z.-M.}\ \bibnamefont {Liao}},\ }\href@noop {} {\bibfield
  {journal} {\bibinfo  {journal} {Advances in Physics: X}\ }\textbf {\bibinfo
  {volume} {2}},\ \bibinfo {pages} {518--544} (\bibinfo {year}
  {2017}{\natexlab{a}})}\BibitemShut {NoStop}%
\bibitem [{\citenamefont {Sattigeri}\ and\ \citenamefont
  {Jha}(2021)}]{14sattigeri2021dimensional}%
  \BibitemOpen
  \bibfield  {author} {\bibinfo {author} {\bibfnamefont {R.~M.}\ \bibnamefont
  {Sattigeri}}\ and\ \bibinfo {author} {\bibfnamefont {P.~K.}\ \bibnamefont
  {Jha}},\ }\href@noop {} {\bibfield  {journal} {\bibinfo  {journal}
  {Scientific Reports}\ }\textbf {\bibinfo {volume} {11}},\ \bibinfo {pages}
  {1--10} (\bibinfo {year} {2021})}\BibitemShut {NoStop}%
\bibitem [{\citenamefont {Graf}, \citenamefont {Felser},\ and\ \citenamefont
  {Parkin}(2011)}]{15graf2011simple}%
  \BibitemOpen
  \bibfield  {author} {\bibinfo {author} {\bibfnamefont {T.}~\bibnamefont
  {Graf}}, \bibinfo {author} {\bibfnamefont {C.}~\bibnamefont {Felser}}, \ and\
  \bibinfo {author} {\bibfnamefont {S.~S.}\ \bibnamefont {Parkin}},\
  }\href@noop {} {\bibfield  {journal} {\bibinfo  {journal} {Progress in solid
  state chemistry}\ }\textbf {\bibinfo {volume} {39}},\ \bibinfo {pages}
  {1--50} (\bibinfo {year} {2011})}\BibitemShut {NoStop}%
\bibitem [{\citenamefont {Casper}\ \emph {et~al.}(2012)\citenamefont {Casper},
  \citenamefont {Graf}, \citenamefont {Chadov}, \citenamefont {Balke},\ and\
  \citenamefont {Felser}}]{16casper2012half}%
  \BibitemOpen
  \bibfield  {author} {\bibinfo {author} {\bibfnamefont {F.}~\bibnamefont
  {Casper}}, \bibinfo {author} {\bibfnamefont {T.}~\bibnamefont {Graf}},
  \bibinfo {author} {\bibfnamefont {S.}~\bibnamefont {Chadov}}, \bibinfo
  {author} {\bibfnamefont {B.}~\bibnamefont {Balke}}, \ and\ \bibinfo {author}
  {\bibfnamefont {C.}~\bibnamefont {Felser}},\ }\href@noop {} {\bibfield
  {journal} {\bibinfo  {journal} {Semiconductor Science and Technology}\
  }\textbf {\bibinfo {volume} {27}},\ \bibinfo {pages} {063001} (\bibinfo
  {year} {2012})}\BibitemShut {NoStop}%
\bibitem [{\citenamefont {Yan}\ and\ \citenamefont
  {de~Visser}(2014)}]{17yan2014half}%
  \BibitemOpen
  \bibfield  {author} {\bibinfo {author} {\bibfnamefont {B.}~\bibnamefont
  {Yan}}\ and\ \bibinfo {author} {\bibfnamefont {A.}~\bibnamefont
  {de~Visser}},\ }\href@noop {} {\bibfield  {journal} {\bibinfo  {journal} {MRS
  Bulletin}\ }\textbf {\bibinfo {volume} {39}},\ \bibinfo {pages} {859--866}
  (\bibinfo {year} {2014})}\BibitemShut {NoStop}%
\bibitem [{\citenamefont {Felser}\ \emph {et~al.}(2015)\citenamefont {Felser},
  \citenamefont {Wollmann}, \citenamefont {Chadov}, \citenamefont {Fecher},\
  and\ \citenamefont {Parkin}}]{18felser2015basics}%
  \BibitemOpen
  \bibfield  {author} {\bibinfo {author} {\bibfnamefont {C.}~\bibnamefont
  {Felser}}, \bibinfo {author} {\bibfnamefont {L.}~\bibnamefont {Wollmann}},
  \bibinfo {author} {\bibfnamefont {S.}~\bibnamefont {Chadov}}, \bibinfo
  {author} {\bibfnamefont {G.~H.}\ \bibnamefont {Fecher}}, \ and\ \bibinfo
  {author} {\bibfnamefont {S.~S.}\ \bibnamefont {Parkin}},\ }\href@noop {}
  {\bibfield  {journal} {\bibinfo  {journal} {APL materials}\ }\textbf
  {\bibinfo {volume} {3}},\ \bibinfo {pages} {041518} (\bibinfo {year}
  {2015})}\BibitemShut {NoStop}%
\bibitem [{\citenamefont {Giannozzi}\ \emph {et~al.}(2009)\citenamefont
  {Giannozzi}, \citenamefont {Baroni}, \citenamefont {Bonini}, \citenamefont
  {Calandra}, \citenamefont {Car}, \citenamefont {Cavazzoni}, \citenamefont
  {Ceresoli}, \citenamefont {Chiarotti}, \citenamefont {Cococcioni},
  \citenamefont {Dabo} \emph {et~al.}}]{19giannozzi2009quantum}%
  \BibitemOpen
  \bibfield  {author} {\bibinfo {author} {\bibfnamefont {P.}~\bibnamefont
  {Giannozzi}}, \bibinfo {author} {\bibfnamefont {S.}~\bibnamefont {Baroni}},
  \bibinfo {author} {\bibfnamefont {N.}~\bibnamefont {Bonini}}, \bibinfo
  {author} {\bibfnamefont {M.}~\bibnamefont {Calandra}}, \bibinfo {author}
  {\bibfnamefont {R.}~\bibnamefont {Car}}, \bibinfo {author} {\bibfnamefont
  {C.}~\bibnamefont {Cavazzoni}}, \bibinfo {author} {\bibfnamefont
  {D.}~\bibnamefont {Ceresoli}}, \bibinfo {author} {\bibfnamefont {G.~L.}\
  \bibnamefont {Chiarotti}}, \bibinfo {author} {\bibfnamefont {M.}~\bibnamefont
  {Cococcioni}}, \bibinfo {author} {\bibfnamefont {I.}~\bibnamefont {Dabo}},
  \emph {et~al.},\ }\href@noop {} {\bibfield  {journal} {\bibinfo  {journal}
  {Journal of physics: Condensed matter}\ }\textbf {\bibinfo {volume} {21}},\
  \bibinfo {pages} {395502} (\bibinfo {year} {2009})}\BibitemShut {NoStop}%
\bibitem [{\citenamefont {Perdew}, \citenamefont {Burke},\ and\ \citenamefont
  {Ernzerhof}(1996)}]{20perdew1996generalized}%
  \BibitemOpen
  \bibfield  {author} {\bibinfo {author} {\bibfnamefont {J.~P.}\ \bibnamefont
  {Perdew}}, \bibinfo {author} {\bibfnamefont {K.}~\bibnamefont {Burke}}, \
  and\ \bibinfo {author} {\bibfnamefont {M.}~\bibnamefont {Ernzerhof}},\
  }\href@noop {} {\bibfield  {journal} {\bibinfo  {journal} {Physical review
  letters}\ }\textbf {\bibinfo {volume} {77}},\ \bibinfo {pages} {3865}
  (\bibinfo {year} {1996})}\BibitemShut {NoStop}%
\bibitem [{\citenamefont {Bl{\"o}chl}(1994)}]{21blochl1994projector}%
  \BibitemOpen
  \bibfield  {author} {\bibinfo {author} {\bibfnamefont {P.~E.}\ \bibnamefont
  {Bl{\"o}chl}},\ }\href@noop {} {\bibfield  {journal} {\bibinfo  {journal}
  {Physical review B}\ }\textbf {\bibinfo {volume} {50}},\ \bibinfo {pages}
  {17953} (\bibinfo {year} {1994})}\BibitemShut {NoStop}%
\bibitem [{\citenamefont {Monkhorst}\ and\ \citenamefont
  {Pack}(1976)}]{22monkhorst1976special}%
  \BibitemOpen
  \bibfield  {author} {\bibinfo {author} {\bibfnamefont {H.~J.}\ \bibnamefont
  {Monkhorst}}\ and\ \bibinfo {author} {\bibfnamefont {J.~D.}\ \bibnamefont
  {Pack}},\ }\href@noop {} {\bibfield  {journal} {\bibinfo  {journal} {Physical
  review B}\ }\textbf {\bibinfo {volume} {13}},\ \bibinfo {pages} {5188}
  (\bibinfo {year} {1976})}\BibitemShut {NoStop}%
\bibitem [{\citenamefont {Grimme}(2006)}]{23grimme2006semiempirical}%
  \BibitemOpen
  \bibfield  {author} {\bibinfo {author} {\bibfnamefont {S.}~\bibnamefont
  {Grimme}},\ }\href@noop {} {\bibfield  {journal} {\bibinfo  {journal}
  {Journal of computational chemistry}\ }\textbf {\bibinfo {volume} {27}},\
  \bibinfo {pages} {1787--1799} (\bibinfo {year} {2006})}\BibitemShut {NoStop}%
\bibitem [{\citenamefont {Wu}\ \emph {et~al.}(2018)\citenamefont {Wu},
  \citenamefont {Zhang}, \citenamefont {Song}, \citenamefont {Troyer},\ and\
  \citenamefont {Soluyanov}}]{24wu2018wanniertools}%
  \BibitemOpen
  \bibfield  {author} {\bibinfo {author} {\bibfnamefont {Q.}~\bibnamefont
  {Wu}}, \bibinfo {author} {\bibfnamefont {S.}~\bibnamefont {Zhang}}, \bibinfo
  {author} {\bibfnamefont {H.-F.}\ \bibnamefont {Song}}, \bibinfo {author}
  {\bibfnamefont {M.}~\bibnamefont {Troyer}}, \ and\ \bibinfo {author}
  {\bibfnamefont {A.~A.}\ \bibnamefont {Soluyanov}},\ }\href@noop {} {\bibfield
   {journal} {\bibinfo  {journal} {Computer Physics Communications}\ }\textbf
  {\bibinfo {volume} {224}},\ \bibinfo {pages} {405--416} (\bibinfo {year}
  {2018})}\BibitemShut {NoStop}%
\bibitem [{\citenamefont {Mostofi}\ \emph {et~al.}(2008)\citenamefont
  {Mostofi}, \citenamefont {Yates}, \citenamefont {Lee}, \citenamefont {Souza},
  \citenamefont {Vanderbilt},\ and\ \citenamefont
  {Marzari}}]{25mostofi2008wannier90}%
  \BibitemOpen
  \bibfield  {author} {\bibinfo {author} {\bibfnamefont {A.~A.}\ \bibnamefont
  {Mostofi}}, \bibinfo {author} {\bibfnamefont {J.~R.}\ \bibnamefont {Yates}},
  \bibinfo {author} {\bibfnamefont {Y.-S.}\ \bibnamefont {Lee}}, \bibinfo
  {author} {\bibfnamefont {I.}~\bibnamefont {Souza}}, \bibinfo {author}
  {\bibfnamefont {D.}~\bibnamefont {Vanderbilt}}, \ and\ \bibinfo {author}
  {\bibfnamefont {N.}~\bibnamefont {Marzari}},\ }\href@noop {} {\bibfield
  {journal} {\bibinfo  {journal} {Computer physics communications}\ }\textbf
  {\bibinfo {volume} {178}},\ \bibinfo {pages} {685--699} (\bibinfo {year}
  {2008})}\BibitemShut {NoStop}%
\bibitem [{\citenamefont {Jayabal}\ \emph {et~al.}(2018)\citenamefont
  {Jayabal}, \citenamefont {Wu}, \citenamefont {Chen}, \citenamefont {Geng},\
  and\ \citenamefont {Meng}}]{26jayabal2018metallic}%
  \BibitemOpen
  \bibfield  {author} {\bibinfo {author} {\bibfnamefont {S.}~\bibnamefont
  {Jayabal}}, \bibinfo {author} {\bibfnamefont {J.}~\bibnamefont {Wu}},
  \bibinfo {author} {\bibfnamefont {J.}~\bibnamefont {Chen}}, \bibinfo {author}
  {\bibfnamefont {D.}~\bibnamefont {Geng}}, \ and\ \bibinfo {author}
  {\bibfnamefont {X.}~\bibnamefont {Meng}},\ }\href@noop {} {\bibfield
  {journal} {\bibinfo  {journal} {Materials today energy}\ }\textbf {\bibinfo
  {volume} {10}},\ \bibinfo {pages} {264--279} (\bibinfo {year}
  {2018})}\BibitemShut {NoStop}%
\bibitem [{\citenamefont {Zhao}\ and\ \citenamefont
  {Wang}(2020)}]{27zhao2020two}%
  \BibitemOpen
  \bibfield  {author} {\bibinfo {author} {\bibfnamefont {A.}~\bibnamefont
  {Zhao}}\ and\ \bibinfo {author} {\bibfnamefont {B.}~\bibnamefont {Wang}},\
  }\href@noop {} {\bibfield  {journal} {\bibinfo  {journal} {APL Materials}\
  }\textbf {\bibinfo {volume} {8}},\ \bibinfo {pages} {030701} (\bibinfo {year}
  {2020})}\BibitemShut {NoStop}%
\bibitem [{\citenamefont {Lee}, \citenamefont {Jun},\ and\ \citenamefont
  {Lee}(2017)}]{28lee2017theoretical}%
  \BibitemOpen
  \bibfield  {author} {\bibinfo {author} {\bibfnamefont {C.~H.}\ \bibnamefont
  {Lee}}, \bibinfo {author} {\bibfnamefont {B.}~\bibnamefont {Jun}}, \ and\
  \bibinfo {author} {\bibfnamefont {S.~U.}\ \bibnamefont {Lee}},\ }\href@noop
  {} {\bibfield  {journal} {\bibinfo  {journal} {RSC advances}\ }\textbf
  {\bibinfo {volume} {7}},\ \bibinfo {pages} {27033--27039} (\bibinfo {year}
  {2017})}\BibitemShut {NoStop}%
\bibitem [{\citenamefont {N{\o}rskov}\ \emph {et~al.}(2005)\citenamefont
  {N{\o}rskov}, \citenamefont {Bligaard}, \citenamefont {Logadottir},
  \citenamefont {Kitchin}, \citenamefont {Chen}, \citenamefont {Pandelov},\
  and\ \citenamefont {Stimming}}]{29norskov2005trends}%
  \BibitemOpen
  \bibfield  {author} {\bibinfo {author} {\bibfnamefont {J.~K.}\ \bibnamefont
  {N{\o}rskov}}, \bibinfo {author} {\bibfnamefont {T.}~\bibnamefont
  {Bligaard}}, \bibinfo {author} {\bibfnamefont {A.}~\bibnamefont
  {Logadottir}}, \bibinfo {author} {\bibfnamefont {J.}~\bibnamefont {Kitchin}},
  \bibinfo {author} {\bibfnamefont {J.~G.}\ \bibnamefont {Chen}}, \bibinfo
  {author} {\bibfnamefont {S.}~\bibnamefont {Pandelov}}, \ and\ \bibinfo
  {author} {\bibfnamefont {U.}~\bibnamefont {Stimming}},\ }\href@noop {}
  {\bibfield  {journal} {\bibinfo  {journal} {Journal of The Electrochemical
  Society}\ }\textbf {\bibinfo {volume} {152}},\ \bibinfo {pages} {J23}
  (\bibinfo {year} {2005})}\BibitemShut {NoStop}%
\bibitem [{\citenamefont {Wang}\ \emph
  {et~al.}(2017{\natexlab{b}})\citenamefont {Wang}, \citenamefont {Zhang},
  \citenamefont {Zhang}, \citenamefont {Wan}, \citenamefont {Guo},
  \citenamefont {Lu}, \citenamefont {Yao},\ and\ \citenamefont
  {Huang}}]{30Pt-wang2017precise}%
  \BibitemOpen
  \bibfield  {author} {\bibinfo {author} {\bibfnamefont {P.}~\bibnamefont
  {Wang}}, \bibinfo {author} {\bibfnamefont {X.}~\bibnamefont {Zhang}},
  \bibinfo {author} {\bibfnamefont {J.}~\bibnamefont {Zhang}}, \bibinfo
  {author} {\bibfnamefont {S.}~\bibnamefont {Wan}}, \bibinfo {author}
  {\bibfnamefont {S.}~\bibnamefont {Guo}}, \bibinfo {author} {\bibfnamefont
  {G.}~\bibnamefont {Lu}}, \bibinfo {author} {\bibfnamefont {J.}~\bibnamefont
  {Yao}}, \ and\ \bibinfo {author} {\bibfnamefont {X.}~\bibnamefont {Huang}},\
  }\href@noop {} {\bibfield  {journal} {\bibinfo  {journal} {Nature
  communications}\ }\textbf {\bibinfo {volume} {8}},\ \bibinfo {pages} {1--9}
  (\bibinfo {year} {2017}{\natexlab{b}})}\BibitemShut {NoStop}%
\bibitem [{\citenamefont {Qu}\ \emph {et~al.}(2021)\citenamefont {Qu},
  \citenamefont {Liu}, \citenamefont {Lau}, \citenamefont {Pan},\ and\
  \citenamefont {Sou}}]{31Pd-SnTe-qu2021highly}%
  \BibitemOpen
  \bibfield  {author} {\bibinfo {author} {\bibfnamefont {Q.}~\bibnamefont
  {Qu}}, \bibinfo {author} {\bibfnamefont {B.}~\bibnamefont {Liu}}, \bibinfo
  {author} {\bibfnamefont {W.~S.}\ \bibnamefont {Lau}}, \bibinfo {author}
  {\bibfnamefont {D.}~\bibnamefont {Pan}}, \ and\ \bibinfo {author}
  {\bibfnamefont {I.~K.}\ \bibnamefont {Sou}},\ }\href@noop {} {\bibfield
  {journal} {\bibinfo  {journal} {arXiv preprint arXiv:2112.04753}\ } (\bibinfo
  {year} {2021})}\BibitemShut {NoStop}%
\bibitem [{\citenamefont {Zhao}\ \emph {et~al.}(2020)\citenamefont {Zhao},
  \citenamefont {Yang}, \citenamefont {Singh}, \citenamefont {Panda},
  \citenamefont {Luo}, \citenamefont {Li},\ and\ \citenamefont
  {Ahuja}}]{32h-B2O-zhao2020strain}%
  \BibitemOpen
  \bibfield  {author} {\bibinfo {author} {\bibfnamefont {X.}~\bibnamefont
  {Zhao}}, \bibinfo {author} {\bibfnamefont {X.}~\bibnamefont {Yang}}, \bibinfo
  {author} {\bibfnamefont {D.}~\bibnamefont {Singh}}, \bibinfo {author}
  {\bibfnamefont {P.~K.}\ \bibnamefont {Panda}}, \bibinfo {author}
  {\bibfnamefont {W.}~\bibnamefont {Luo}}, \bibinfo {author} {\bibfnamefont
  {Y.}~\bibnamefont {Li}}, \ and\ \bibinfo {author} {\bibfnamefont
  {R.}~\bibnamefont {Ahuja}},\ }\href@noop {} {\bibfield  {journal} {\bibinfo
  {journal} {The Journal of Physical Chemistry C}\ }\textbf {\bibinfo {volume}
  {124}},\ \bibinfo {pages} {7884--7892} (\bibinfo {year} {2020})}\BibitemShut
  {NoStop}%
\bibitem [{\citenamefont {Liu}\ \emph {et~al.}(2022)\citenamefont {Liu},
  \citenamefont {Zhang}, \citenamefont {Meng}, \citenamefont {Liu},
  \citenamefont {Dai},\ and\ \citenamefont {Liu}}]{33NiSi-liu2022theoretical}%
  \BibitemOpen
  \bibfield  {author} {\bibinfo {author} {\bibfnamefont {W.}~\bibnamefont
  {Liu}}, \bibinfo {author} {\bibfnamefont {X.}~\bibnamefont {Zhang}}, \bibinfo
  {author} {\bibfnamefont {W.}~\bibnamefont {Meng}}, \bibinfo {author}
  {\bibfnamefont {Y.}~\bibnamefont {Liu}}, \bibinfo {author} {\bibfnamefont
  {X.}~\bibnamefont {Dai}}, \ and\ \bibinfo {author} {\bibfnamefont
  {G.}~\bibnamefont {Liu}},\ }\href@noop {} {\bibfield  {journal} {\bibinfo
  {journal} {Iscience}\ }\textbf {\bibinfo {volume} {25}},\ \bibinfo {pages}
  {103543} (\bibinfo {year} {2022})}\BibitemShut {NoStop}%
\bibitem [{\citenamefont {Hinnemann}\ \emph {et~al.}(2005)\citenamefont
  {Hinnemann}, \citenamefont {Moses}, \citenamefont {Bonde}, \citenamefont
  {J{\o}rgensen}, \citenamefont {Nielsen}, \citenamefont {Horch}, \citenamefont
  {Chorkendorff},\ and\ \citenamefont
  {N{\o}rskov}}]{34MoS2-hinnemann2005biomimetic}%
  \BibitemOpen
  \bibfield  {author} {\bibinfo {author} {\bibfnamefont {B.}~\bibnamefont
  {Hinnemann}}, \bibinfo {author} {\bibfnamefont {P.~G.}\ \bibnamefont
  {Moses}}, \bibinfo {author} {\bibfnamefont {J.}~\bibnamefont {Bonde}},
  \bibinfo {author} {\bibfnamefont {K.~P.}\ \bibnamefont {J{\o}rgensen}},
  \bibinfo {author} {\bibfnamefont {J.~H.}\ \bibnamefont {Nielsen}}, \bibinfo
  {author} {\bibfnamefont {S.}~\bibnamefont {Horch}}, \bibinfo {author}
  {\bibfnamefont {I.}~\bibnamefont {Chorkendorff}}, \ and\ \bibinfo {author}
  {\bibfnamefont {J.~K.}\ \bibnamefont {N{\o}rskov}},\ }\href@noop {}
  {\bibfield  {journal} {\bibinfo  {journal} {Journal of the American Chemical
  Society}\ }\textbf {\bibinfo {volume} {127}},\ \bibinfo {pages} {5308--5309}
  (\bibinfo {year} {2005})}\BibitemShut {NoStop}%
\bibitem [{\citenamefont {Li}\ \emph {et~al.}(2018)\citenamefont {Li},
  \citenamefont {Ma}, \citenamefont {Xie}, \citenamefont {Feng}, \citenamefont
  {Ullah}, \citenamefont {Li}, \citenamefont {Dong}, \citenamefont {Li},
  \citenamefont {Li},\ and\ \citenamefont {Chen}}]{35TiSi-li2018topological}%
  \BibitemOpen
  \bibfield  {author} {\bibinfo {author} {\bibfnamefont {J.}~\bibnamefont
  {Li}}, \bibinfo {author} {\bibfnamefont {H.}~\bibnamefont {Ma}}, \bibinfo
  {author} {\bibfnamefont {Q.}~\bibnamefont {Xie}}, \bibinfo {author}
  {\bibfnamefont {S.}~\bibnamefont {Feng}}, \bibinfo {author} {\bibfnamefont
  {S.}~\bibnamefont {Ullah}}, \bibinfo {author} {\bibfnamefont
  {R.}~\bibnamefont {Li}}, \bibinfo {author} {\bibfnamefont {J.}~\bibnamefont
  {Dong}}, \bibinfo {author} {\bibfnamefont {D.}~\bibnamefont {Li}}, \bibinfo
  {author} {\bibfnamefont {Y.}~\bibnamefont {Li}}, \ and\ \bibinfo {author}
  {\bibfnamefont {X.-Q.}\ \bibnamefont {Chen}},\ }\href@noop {} {\bibfield
  {journal} {\bibinfo  {journal} {Science China Materials}\ }\textbf {\bibinfo
  {volume} {61}},\ \bibinfo {pages} {23--29} (\bibinfo {year}
  {2018})}\BibitemShut {NoStop}%
\bibitem [{\citenamefont {He}\ \emph {et~al.}(2019)\citenamefont {He},
  \citenamefont {Yan}, \citenamefont {Ng}, \citenamefont {Shi}, \citenamefont
  {Wang}, \citenamefont {Lin}, \citenamefont {Lin}, \citenamefont {Luo},\ and\
  \citenamefont {Yan}}]{36BiPd-he2019topological}%
  \BibitemOpen
  \bibfield  {author} {\bibinfo {author} {\bibfnamefont {Y.}~\bibnamefont
  {He}}, \bibinfo {author} {\bibfnamefont {D.}~\bibnamefont {Yan}}, \bibinfo
  {author} {\bibfnamefont {L.~R.}\ \bibnamefont {Ng}}, \bibinfo {author}
  {\bibfnamefont {L.}~\bibnamefont {Shi}}, \bibinfo {author} {\bibfnamefont
  {S.}~\bibnamefont {Wang}}, \bibinfo {author} {\bibfnamefont {H.}~\bibnamefont
  {Lin}}, \bibinfo {author} {\bibfnamefont {S.-H.}\ \bibnamefont {Lin}},
  \bibinfo {author} {\bibfnamefont {H.}~\bibnamefont {Luo}}, \ and\ \bibinfo
  {author} {\bibfnamefont {K.}~\bibnamefont {Yan}},\ }\href@noop {} {\bibfield
  {journal} {\bibinfo  {journal} {Materials Chemistry Frontiers}\ }\textbf
  {\bibinfo {volume} {3}},\ \bibinfo {pages} {2184--2189} (\bibinfo {year}
  {2019})}\BibitemShut {NoStop}%
\bibitem [{\citenamefont {Kong}\ \emph {et~al.}(2021)\citenamefont {Kong},
  \citenamefont {Jiang}, \citenamefont {Gao}, \citenamefont {Shi},
  \citenamefont {Shao}, \citenamefont {Yuan}, \citenamefont {Qiu},\ and\
  \citenamefont {Zhao}}]{37Ni-VAl3-kong2021development}%
  \BibitemOpen
  \bibfield  {author} {\bibinfo {author} {\bibfnamefont {X.-P.}\ \bibnamefont
  {Kong}}, \bibinfo {author} {\bibfnamefont {T.}~\bibnamefont {Jiang}},
  \bibinfo {author} {\bibfnamefont {J.}~\bibnamefont {Gao}}, \bibinfo {author}
  {\bibfnamefont {X.}~\bibnamefont {Shi}}, \bibinfo {author} {\bibfnamefont
  {J.}~\bibnamefont {Shao}}, \bibinfo {author} {\bibfnamefont {Y.}~\bibnamefont
  {Yuan}}, \bibinfo {author} {\bibfnamefont {H.-J.}\ \bibnamefont {Qiu}}, \
  and\ \bibinfo {author} {\bibfnamefont {W.}~\bibnamefont {Zhao}},\ }\href@noop
  {} {\bibfield  {journal} {\bibinfo  {journal} {The Journal of Physical
  Chemistry Letters}\ }\textbf {\bibinfo {volume} {12}},\ \bibinfo {pages}
  {3740--3748} (\bibinfo {year} {2021})}\BibitemShut {NoStop}%
\bibitem [{\citenamefont {Yang}\ \emph {et~al.}(2021)\citenamefont {Yang},
  \citenamefont {Li}, \citenamefont {Zhang}, \citenamefont {Liu}, \citenamefont
  {Rao}, \citenamefont {Heine}, \citenamefont {Felser},\ and\ \citenamefont
  {Sun}}]{38W-PdGa-yang2021transition}%
  \BibitemOpen
  \bibfield  {author} {\bibinfo {author} {\bibfnamefont {Q.}~\bibnamefont
  {Yang}}, \bibinfo {author} {\bibfnamefont {G.}~\bibnamefont {Li}}, \bibinfo
  {author} {\bibfnamefont {Y.}~\bibnamefont {Zhang}}, \bibinfo {author}
  {\bibfnamefont {J.}~\bibnamefont {Liu}}, \bibinfo {author} {\bibfnamefont
  {J.}~\bibnamefont {Rao}}, \bibinfo {author} {\bibfnamefont {T.}~\bibnamefont
  {Heine}}, \bibinfo {author} {\bibfnamefont {C.}~\bibnamefont {Felser}}, \
  and\ \bibinfo {author} {\bibfnamefont {Y.}~\bibnamefont {Sun}},\ }\href@noop
  {} {\bibfield  {journal} {\bibinfo  {journal} {npj Computational Materials}\
  }\textbf {\bibinfo {volume} {7}},\ \bibinfo {pages} {1--8} (\bibinfo {year}
  {2021})}\BibitemShut {NoStop}%
\bibitem [{\citenamefont {Ekspong}, \citenamefont {Gracia-Espino},\ and\
  \citenamefont {Wagberg}(2020)}]{39vol-plot2-ekspong2020hydrogen}%
  \BibitemOpen
  \bibfield  {author} {\bibinfo {author} {\bibfnamefont {J.}~\bibnamefont
  {Ekspong}}, \bibinfo {author} {\bibfnamefont {E.}~\bibnamefont
  {Gracia-Espino}}, \ and\ \bibinfo {author} {\bibfnamefont {T.}~\bibnamefont
  {Wagberg}},\ }\href@noop {} {\bibfield  {journal} {\bibinfo  {journal} {The
  Journal of Physical Chemistry C}\ }\textbf {\bibinfo {volume} {124}},\
  \bibinfo {pages} {20911--20921} (\bibinfo {year} {2020})}\BibitemShut
  {NoStop}%
\bibitem [{\citenamefont {Yang}\ \emph {et~al.}(2020)\citenamefont {Yang},
  \citenamefont {Li}, \citenamefont {Manna}, \citenamefont {Fan}, \citenamefont
  {Felser},\ and\ \citenamefont {Sun}}]{40PtGa-PtAl-yang2020topological}%
  \BibitemOpen
  \bibfield  {author} {\bibinfo {author} {\bibfnamefont {Q.}~\bibnamefont
  {Yang}}, \bibinfo {author} {\bibfnamefont {G.}~\bibnamefont {Li}}, \bibinfo
  {author} {\bibfnamefont {K.}~\bibnamefont {Manna}}, \bibinfo {author}
  {\bibfnamefont {F.}~\bibnamefont {Fan}}, \bibinfo {author} {\bibfnamefont
  {C.}~\bibnamefont {Felser}}, \ and\ \bibinfo {author} {\bibfnamefont
  {Y.}~\bibnamefont {Sun}},\ }\href@noop {} {\bibfield  {journal} {\bibinfo
  {journal} {Advanced Materials}\ }\textbf {\bibinfo {volume} {32}},\ \bibinfo
  {pages} {1908518} (\bibinfo {year} {2020})}\BibitemShut {NoStop}%
\bibitem [{\citenamefont {Wang}\ \emph {et~al.}(2021)\citenamefont {Wang},
  \citenamefont {Zhang}, \citenamefont {Meng}, \citenamefont {Liu},
  \citenamefont {Dai},\ and\ \citenamefont
  {Liu}}]{41Cu2C2N4-wang2021topological}%
  \BibitemOpen
  \bibfield  {author} {\bibinfo {author} {\bibfnamefont {L.}~\bibnamefont
  {Wang}}, \bibinfo {author} {\bibfnamefont {X.}~\bibnamefont {Zhang}},
  \bibinfo {author} {\bibfnamefont {W.}~\bibnamefont {Meng}}, \bibinfo {author}
  {\bibfnamefont {Y.}~\bibnamefont {Liu}}, \bibinfo {author} {\bibfnamefont
  {X.}~\bibnamefont {Dai}}, \ and\ \bibinfo {author} {\bibfnamefont
  {G.}~\bibnamefont {Liu}},\ }\href@noop {} {\bibfield  {journal} {\bibinfo
  {journal} {Journal of Materials Chemistry A}\ }\textbf {\bibinfo {volume}
  {9}},\ \bibinfo {pages} {22453--22461} (\bibinfo {year} {2021})}\BibitemShut
  {NoStop}%
\end{thebibliography}%

\end{document}